\documentclass[final,2p,times]{elsarticle}
\usepackage{graphicx}
\usepackage{bbm}
\usepackage{amsfonts,bbm,amsfonts}
\usepackage[OT4,T1]{fontenc}
\usepackage[cp1250]{inputenc}
\usepackage{mathrsfs}
\usepackage{textcomp}

\biboptions{comma,square,numbers,sort&compress}
\journal{Physics Letters A}

\begin{document}

\begin{frontmatter}

\title{Squeezed coherent state undergoing a continuous
nondemolition observation}

\author{Anita D\c{a}browska}
\ead{adabro@cm.umk.pl}
\author{Przemysław Staszewski}
\ead{przemek.staszewski@cm.umk.pl}

\address{Department of Theoretical Foundations of Biomedical
Sciences and
Medical Informatics\\
Collegium Medicum, Nicolaus Copernicus University\\
ul.~Jagiello\'nska 15, 85-067 Bydgoszcz, Poland}

\begin{abstract}
The time evolution of a squeezed coherent state conditioned by the
results of a single and double heterodyne measurement is
discussed. The mean values of quadratures as well as the dynamics
of quadrature uncertainties have been obtained within the
framework of the theory of continuous measurements based on
filtration equations. It has been found that while the mean values
depend on the measured noise, the uncertainties in the optical
quadratures are deterministic. Explicit solutions for the latter
have been provided. Finally, a time development of the squeeze
parameter for the posterior squeezed coherent state has been
found.
\end{abstract}
\end{frontmatter}

\section{Introduction}

The problem of continuous measurements in quantum systems is one
of the most challenging fundamental issues of modern theoretical
physics \cite{GZ2000, BG2009}. The quantum filtering theory
developed by Belavkin \cite{Bel89,Bel90,BarBel91} makes it
possible to describe the dynamics of a quantum system continuously
observed in time. The r\^ole of a measuring apparatus is played
here by a Bose field which in quantum optics can be treated as an
approximation to the electromagnetic field. The interaction
between the quantum system in question (system $\mathcal{S}$) and
the reservoir (the Bose field) is taken in the Markovian
approximation, i.e. the correlation time of the reservoir is much
shorter than the time scale of the dynamics of S. The continuous
trajectory of results of the observation of the output Bose field
(Bose field after interaction with $\mathcal{S}$) determines the
conditioned evolution of $\mathcal{S}$ (the posterior state). The
main motivation for determining the conditional state of a quantum
system is developing  methods of the quantum feedback control. The
quantum trajectory has been essential to the design of a quantum
control algorithm \cite{WM2010}.

In this article, we present the posterior evolution of an optical
cavity mode  coupled to the outside radiation mode by one or two
partially transmitting  mirrors. The time-development of a state
of an indirectly observed system is conditioned by a trajectory of
the results of a single and double heterodyne measurement. We
assume the coupling system's operator proportional to the
annihilation operator. In the paper we discuss the analytical
solutions to the filtering equation for the system being initially
in a squeezed coherent state. In order to prove that such a state
is preserved under the considered continuous diffusion observation
we use the linear version of the quantum filtering equation
derived in \cite{Bel90}.

The exact solutions to the filtering equation for the initial
Gaussian states and a diffusion observation were given, for
instance, in \cite{BelSta89, BarBel91, BelSta92, StaSta00,
StaSta01}. The solution to the quantum filtering equation for a
harmonic oscillator in an arbitrary initial state state undergoing
the heterodyne observation was discussed by Carmichael in
\cite{Car08}.

The main part of our work is organized as follows. In Section~2
the mathematical model is described. Section~3 contains the
analysis of the double heterodyne detection problem. Section~4 is
devoted to the single heterodyne detection. Some final remarks are
given in Section~5.

\section{Model}

We consider a single cavity mode of the electromagnetic field (a
system $\mathcal{S}$) interacting with two independent components
$B_{n}(t)\,,(n=1,2)$ of the Bose field being initially in the
vacuum state. The unitary operator, $U(t)$, describing the
evolution of the whole system (the system $\mathcal{S}$ plus Bose
field) satisfies the Ito quantum stochastic differential equation
(QSDE):
\begin{eqnarray}\label{unitary}
\lefteqn{\mathrm{d}U(t)\;=\; \bigg[-\left(\frac{\mathrm{i}}{\hbar}
H+\sum_{n=1}^{2}{\mu_{n} \over 4}a^{\dagger}a\right)\mathrm{d}t }
\nonumber \\
&&+ \sum_{n=1}^{2} \sqrt{\frac{\mu_{n}}{2}}\bigg(a\
\mathrm{d}B_{n}^{\dagger} (t)-a^{\dagger}\
\mathrm{d}B_{n}(t)\bigg)\bigg]U(t)\,,\nonumber
\\&& U(0)\;=\;I\,,
\end{eqnarray}
where  $H=\hbar \omega\left(a^{\dagger}a+{1\over 2}\right)$ is the
hamiltonian of $\mathcal{S}$, $a$ stands for the
annihilation operator of $\mathcal{S}$ and $\mu_{n} > 0\, (n=1,2)$
are coupling constants. Eq.~(\ref{unitary}) is written
in the interaction picture with respect to the free dynamics of the Bose
field. The discussion of the physical
assumptions leading to this evolution one can find, for instance,
in \cite{GarCol84, Bar06}. In brief, the interaction
hamiltonian is taken linear in the field operators, the rotating-wave
approximation (RWA) is made and a flat and broad
spectrum of the reservoir is assumed.

The Bose field provides a possibility of a continuous indirect
observation of $\mathcal{S}$. We suppose that the information
about the systems is gained by using a double heterodyne detection
scheme, thus we consider a simultaneous measurement of the two
output processes
\begin{eqnarray}\label{observation}
\lefteqn{\mathcal{Q}^{\mathrm{out}}_{n}(t)\;=\;
\int\limits_{0}^{t}\big(
\mathrm{e}^{\mathrm{i}\phi_{n}(t^{\prime})} \mathrm{d}
B_{n}^{\dagger}(t^{\prime})+\mathrm{e}^{-\mathrm{i}\phi_{n}(t^{\prime})}
\mathrm{d}B_{n}(t^{\prime})\big)} \nonumber \\
&&+\sqrt{\mu_{n}/2}\, \mathrm{Re}\big( \mathrm{e}^{-\mathrm{i}\phi_{n}
(t^{\prime})}a\big)
\mathrm{d}t^{\prime}\,,\;\;\;\;n=1,2\,,
\end{eqnarray}
where $\phi(t)=\phi_{0}+\vartheta t$ \cite{CLG87, Bar90}. These output
processes satisfy the Belavkin's nondemolition
condition, namely
\begin{equation}
[\mathcal{Q}^{\mathrm{out}}_{n}(s),U^{\dagger}(t)ZU(t)]\;=\;0\,\;\;\;\;\;
\forall s\leq t\,,
\end{equation}
where $Z$ is any operator of the system $\mathcal{S}$.
Furthermore, the output processes (\ref{observation}), due to
their Hermicity and self-commutativity,
\begin{equation}
[\mathcal{Q}^{\mathrm{out}}_{n}(t),\mathcal{Q}^{\mathrm{out}}_{n}
(t^{\prime})] \;=\;0\,\;\;\; \forall t\,,t^{\prime}\geq 0\,,
\end{equation}
can be treated as the classical Wiener processes.

The time development of the posterior unnormalized wave function
$\widehat{\psi}(t)$ of $\mathcal{S}$, corresponding to the
trajectory of the observed processes (\ref{observation}) up to
$t$, is given by the Belavkin linear filtering equation of the
form
\begin{eqnarray}\label{filequ}
\lefteqn{\mathrm{d}\widehat{\psi}(t)\;=\; - \,
\bigg(\frac{\mathrm{i}}{\hbar}H+
\sum_{n=1}^{2}\frac{\mu_{n}}{4}a^{\dagger}a\bigg)
\,\widehat{\psi}(t)\, \mathrm{d}t} \nonumber \\
&&+\sum_{n=1}^{2}\sqrt{\frac{\mu_{n}}{2}}\, a\,
\mathrm{e}^{-\mathrm{i}\phi_{n}(t)}\, \widehat{\psi}(t) \,
\mathrm{d}\mathcal{Q}_{n}(t) \,,\quad \widehat{\psi}(0)=\psi\,,
\end{eqnarray}
where $\mathcal{Q}_{n}(t)$ ($n=1,2$) are Wiener processes for
which $\mathrm{d}\mathcal{Q}_{n}(t)
\mathrm{d}\mathcal{Q}_{m}(t)=\delta_{nm} \mathrm{d}t$. Let us
recall  that the posterior mean value of any operator $Z$ of the
system $\mathcal{S}$ reads in terms of unnormalized wave function
$\widehat{\psi}(t)$ satisfying the linear filtering equation
\begin{equation}\label{def1}
\langle Z\rangle_{t} \;=\; {\langle \widehat{\psi} (t)|Z
\widehat{\psi}(t) \rangle \over \langle
\widehat{\psi}(t)|\widehat{\psi}(t)\rangle}\,.
\end{equation}
For more details on the r\^ole of the linear filtering equation for the
diffusion observation in quantum mechanics, its derivation, and
some exemplary solutions see for instance \cite{Bel90, GoeGra94}.

\section{Double heterodyne detection}

Let us put: $\mu_{n}=\mu$ ($n=1,2$),
$\phi_{2}(t)-\phi_{1}(t)=\pi/2$, and $\phi_{1}(t)=\phi(t)$. With
these assumptions, the linear filtering equation for the double
heterodyne measurement takes the form
\begin{eqnarray}\label{filtequ1}
&& \!\!\!\!\!\!\!\!\!\!\! \mathrm{d}\widehat{\psi}(t)\;=\;
-\,\bigg(\frac{\mathrm{i}}{\hbar}H+
\frac{\mu}{2}a^{\dagger}a\bigg)\widehat{\psi}(t) \mathrm{d}t+
\sqrt{\mu}\, a\,\mathrm{e}^{-\mathrm{i}\phi(t)}
\widehat{\psi}(t)\, \mathrm{d}\mathcal{Q}(t)\,, \nonumber
\\ && \!\!\!\!\!\!\!\!\!\!\!
\widehat{\psi}(0)=\psi\,,
\end{eqnarray}
where $\mathrm{d}\mathcal{Q}(t)\;=\;\frac{1}{\sqrt{2}}
\left(\mathrm{d}\mathcal{Q}_{1}(t)-\mathrm{i}\,
\mathrm{d}\mathcal{Q}_{2}(t)\right)$ is the complex dif\-fus\-ion
process such that $\big(\mathrm{d}\mathcal{Q}(t)\big)^{2}=0\,$
 and \mbox{$ |\kern1pt \mathrm{d}\mathcal{Q}^{\dagger}(t)\kern.5pt|^2
=\mathrm{d}t\,$.}

Let us discuss the time development of the posterior wave
function, assuming that the initial state of $\mathcal{S}$ is a
squeezed coherent state \cite{MeySar07},
\begin{equation}\label{squ}
\widehat \psi (0) \;=\;S(\xi_{0} )D(\alpha_{0})|0 \rangle\;=\;
S(\xi_{0} )|\alpha_{0} \rangle\;=\;|\xi_{0} ,\alpha_{0}
\rangle\,,
\end{equation}
where
\begin{equation}
D(\alpha_{0})\;=\;\exp\big(\alpha_{0}
a^{\dagger}-\overline{\alpha}_{0} a\big)\,,
\end{equation}
and
\begin{equation}
\!\,S(\xi_{0})= \exp \left( {1\over
2}\overline{\xi}_{0}a^{2}-{1\over 2} \xi_{0}
\big(a^{\dagger}\big)^{2}\right), \,\,\,
\xi_{0}=\mathrm{e}^{\mathrm{i}\theta_{0}} \varrho_{0} \in
\mathbb{C}.
\end{equation}
The amount of squeeze is described by the modulus $\rho_{0}$ of
the squeeze parameter $\xi_0$, whereas the phase $\theta_{0}$
specifies the angle of the squeeze axis in the phase space.
Obviously, for $\xi_{0}=0$ the state (\ref{squ}) becomes the
coherent state $|\alpha_{0}\rangle$, and for $\alpha_{0}=0$ and
$\xi_{0}\neq 0$ it takes the form of the squeezed vacuum state,
$S(\xi_{0})|0 \rangle$. Let us recall that the expectation values
of the quadratures vanish while the expectation value of the
photon number operator is nonzero in the state $S(\xi_{0})|0
\rangle$.

We shall prove that the solution to Eg.~(\ref{filtequ1})
corresponding to the initial state (\ref{squ}) can be written as
\begin{equation}\label{squ2}
\widehat\psi(t)\;=\;l(t)S(\xi(t))|\alpha (t) \rangle\,.
\end{equation}
The proof is rather standard but still quite cumbersome, therefore
we shall present its outline augmented with some computational
details and partial results.

Substituting the postulated solution (\ref{squ2}) into
Eg.~(\ref{filtequ1}), writing  both sides of the equation in terms
of linearly independent vectors $\left\{|\alpha\rangle,
\frac{\partial|\alpha\rangle }{\partial \alpha},
\frac{\partial^{2}| \alpha\rangle}{\partial \alpha^{2}}\right\}$
and comparing the coefficients of the corresponding vectors yield
the consistent system of differential equations for the functions:
$\varrho(t)$, $\theta(t)$, $\alpha(t)$, $l(t)$. The solution to
this system corresponding to the initial state (\ref{squ})
uniquely defines the solution to Eg.~(\ref{filtequ1}) in the form
(\ref{squ2}). Let us pay attention to several steps of these
calculations.

To calculate the increment $\mathrm{d}\widehat{\psi}(t)$ one has
to calculate the increment $\mathrm{d}S\left(\xi(t)\right)=S\left(
\xi(t+\mathrm{d}t)\right)-S(\xi(t))$. To this end it is convenient
to make use of the normally ordered form \cite{WalMil94} of the
squeeze operator, $S(\xi)$,
\begin{eqnarray}
S(\xi)\;=\;\left(\cosh \varrho \right)^{-1/2}\exp\left[-\Gamma\,
\big(a^{\dagger}\big)^{2}/2\right] \nonumber \\
\times \exp \left[-\ln \left(\cosh \varrho
\right)a^{\dagger}a\right]\exp\left[\,\overline{\Gamma}a^{2}/2\right]\,,
\end{eqnarray}
where
\begin{equation}\label{gamma}
\Gamma\;=\;\mathrm{e}^{\mathrm{i}\theta }\tanh \varrho\,.
\end{equation}
Left-multiplying Eq.~(\ref{filtequ1}) by $S^{\dagger}(\xi(t))$ and
using the unitary transformation
\begin{eqnarray}
&&\!\!\!\!\!\!\!\!\!\!\!\!\!\!\!\!\!\!\!\!\!\!S^{\dagger}\left(\xi
\right)\big(a^{\dagger}\big)^{2}S\left(\xi
\right)\;=\;\big(a^{\dagger}\big)^{2}\cosh ^{2}\varrho \nonumber
\\ &&\!\!\!\!\!\!\!\!\!\!\!\!\!\!\!\!\!\! -\,\left( 2a^{\dagger}a+
1\right)\mathrm{e}^{-\mathrm{i}
\theta } \sinh \varrho \cosh \varrho
+a^{2}\mathrm{e}^{-2\mathrm{i}\theta
}\sinh ^{2}\varrho\,,
\end{eqnarray}
one gets
\begin{eqnarray}
&& \!\!\!\!\!\!\!\!\!\!\!
S^{\dagger}(\xi(t))\mathrm{d}S(\xi(t))=\frac{1}{2}\tanh \varrho(t)
\mathrm{d}\varrho(t)\left[2\overline{\Gamma}(t)a^{2}-2a^{\dagger}a-1
\right]\nonumber
\\ && \!\!\!\! +\frac{\mathrm{d}\overline{\Gamma}(t)}{2}\,a^{2} -
\frac{\mathrm{d}\Gamma(t)}{2} \left[
\big(a^{\dagger}\big)^{2}\cosh ^{2}\varrho(t)\right.\\ &&
\!\!\!\!\!\!\!\!\! -\!\left.\left(2a^{\dagger}a\!+\!1
\!\right)\mathrm{e}^{-\mathrm{i}\theta(t)
}\sinh \varrho(t) \cosh \varrho(t)
\!+\!a^{2}\mathrm{e}^{-2\mathrm{i}\theta(t) }\sinh
^{2}\varrho(t)\right]\!.\nonumber
\end{eqnarray}
By the virtue of applying the formulae
\begin{eqnarray}
\!\!\!\!\!\!\!\!\!\!\!\!S^{\dagger}(\xi )aS(\xi )&=& a\cosh
\varrho -a^{\dagger}
\mathrm{e}^{\mathrm{i}\theta }\sinh \varrho\,, \\
\!\!\!\!\!\!\!\!\!\!\!\!S^{\dagger}(\xi )a^{\dagger}aS(\xi ) &=&
a^{\dagger}a\cosh ^{2}\varrho
+\left(a^{\dagger}a+1\right)\sinh^{2}\varrho \nonumber
\\&& \!\!\!\!- \left[a^{2}\mathrm{e}^{-\mathrm{i}\theta
}+\big(a^{\dagger}\big)^{2}\mathrm{e}^{\mathrm{i}\theta
}\right]\sinh \varrho \cosh \varrho
\end{eqnarray}
and
\begin{eqnarray}\label{ccpro}
a^{\dagger}|\alpha\rangle&=&\frac{\partial|\alpha\rangle}{\partial
\alpha}+\frac{1}{2}\overline{\alpha}|\alpha\rangle\,,\\
(a^{\dagger})^{2}|\alpha\rangle &=&
\frac{\partial^{2}|\alpha\rangle}
{\partial \alpha^{2}}+\overline{\alpha}\frac{\partial|\alpha\rangle}
{\partial\alpha} +\frac{1}{4}
\overline{\alpha}^{2}|\alpha\rangle\,,
\end{eqnarray}
one is able to rewrite the equation in terms of the vectors
$\left\{|\alpha\rangle, \frac{\partial|\alpha\rangle
}{\partial \alpha}, \frac{\partial^{2}| \alpha\rangle}{\partial
\alpha^{2}}\right\}$. As these vectors are linearly independent,
the comparison of the expansion coefficients yields the set of
differential equations for the functions $\theta$, $\varrho$,
$\alpha$, and $l$:
\begin{eqnarray}\label{sys2}
\mathrm{d}\theta (t)&\!=\!&-2\omega \mathrm{d}t\,,\;\;\;
\mathrm{d} \varrho (t)\;=\;-\mu \sinh \varrho (t)\cosh \varrho
(t)\mathrm{d}t\,,\nonumber\\
\mathrm{d}\alpha(t)&\!=\!&\left[-\left(\mathrm{i}\omega +
{\mu \over 2} \right)-\mu \sinh^{2}\varrho (t)\right]\alpha
(t)\mathrm{d}t \nonumber \\
&&-\sqrt {\mu }\,
\mathrm{e}^{\mathrm{i}\theta (t)}\sinh \varrho
(t)\mathrm{e}^{-\mathrm{i}\phi(t)}\mathrm{d} \mathcal{Q}(t)\,,\\
\frac{\mathrm{d}l(t)}{l(t)}&=&-{\mathrm{i}\omega \over 2} \mathrm{d}t+
{1\over 2} \mathrm{d}\left|\alpha (t)\right|^{2}+
{\mu \over 4} \sinh ^{2}\varrho (t)\left|\alpha(t)\right|^{2}
\mathrm{d}t \nonumber \\
&& + \sqrt {\mu }\alpha (t)\cosh
\varrho (t) \mathrm{e}^{-\mathrm{i}\phi(t)}\mathrm{d}
\mathcal{Q}(t)-{\mu \over 2} \sinh ^{2}\varrho (t)\mathrm{d}t\nonumber\\
&&+{\mu \over 2} \alpha^{2}(t) \mathrm{e}^{-\mathrm{i}
\theta (t)} \sinh \varrho (t) \cosh
\varrho(t)\mathrm{d}t\,,\nonumber
\end{eqnarray}
with the initial condition $l(0) = 1$, $\alpha(0) =
\alpha_{0}$, $\theta(0) = \theta_{0}$, $\varrho(0) = \varrho_{0}$.

\noindent The solution to the system (\ref{sys2}) reads
\begin{eqnarray}\label{sol2}
\theta (t)&\!\!\!\!\!\!=\!\!\!\!\!\!&\theta_{0} -2\omega t\,, \;\;\;\;
\varrho (t)=\mathrm{ar}\tanh
\left(\mathrm{e}^{-\mu t}\tanh \varrho_{0} \right)\,,\nonumber \\
\alpha (t)&\!\!\!\!\!\!=\!\!\!\!\!\!&
\mathrm{e}^{-\left(\mathrm{i}\omega +{\mu \over 2}
\right)t}\,{\cosh \varrho (t)\over \cosh
\varrho_{0} } \bigg( \alpha_{0}  \nonumber \\
&&  -\sqrt {\mu }\, \mathrm{e}^{\mathrm{i}\theta_{0}
} \sinh \varrho_{0} \int\limits
_{0}^{t}\mathrm{e}^{-\left(\!\!\mathrm{i}\omega +{\mu \over 2}
\right)t^{\prime}}\mathrm{e}^{-\mathrm{i}\phi(t^{\prime})}
\mathrm{d} \mathcal{Q}(t^{\prime})\! \bigg),\\
l(t)&\!\!\!\!\!\!\!\!\!=\!\!\!\!\!\!\!\!\!&\sqrt {\cosh \varrho
(t)\over \cosh \varrho_{0}
}\! \exp\bigg[\!\!-{\mathrm{i}\omega t\over 2}\!+\!{1\over
2}\left(\left| \alpha (t)
\right|^{2}-\left|\alpha_{0}\right|^{2}\right)\!+\!\chi(t)\!\bigg],
\nonumber
\end{eqnarray}
where
\begin{eqnarray*}
\chi(t)&=& \sqrt {\mu}\!\int\limits
_{0}^{t}\left(\alpha(t^{\prime}) \cosh \varrho
(t^{\prime})\mathrm{e}^{-\mathrm{i}\phi(t^{\prime})}
\mathrm{d}\mathcal{Q}(t^{\prime}) \right. \\
&&\left. + \sqrt{\mu} \mathrm{e}^{-\mathrm{i}\theta
(t^{\prime})}\alpha ^{2}(t^{\prime}) \sinh \varrho (t^{\prime})
\cosh \varrho(t^{\prime})\mathrm{d}t^{\prime}\right)\,.
\end{eqnarray*}
Therefore the posterior mean values of optical quadratures
$X=(a+a^{\dagger})/2$ and $Y=(a-a^{\dagger})/2\mathrm{i}$ for the
posterior wave function of the form (\ref{squ2}), given by
\begin{equation}
\langle X\rangle_{t}\;=\;\mathrm{Re} \left(\alpha
\left(t\right)\cosh \varrho (t)-\overline{\alpha}
\left(t\right)\,\mathrm{e}^{\mathrm{i}\theta (t)}\sinh \varrho (t)
\right)\,,
\end{equation}
\begin{equation}
\langle Y\rangle_{t}\;=\;\mathrm{Im} \left(\alpha (t)\cosh \varrho (t)+
\alpha (t)\mathrm{e}^{-\mathrm{i}\theta
(t)} \sinh \varrho (t) \right)\,,
\end{equation}
depend on the measured noise, whereas the uncertainties in the
optical quadratures $\triangle X(t)$ and $\triangle Y(t)$ are
deterministic. One has:
\begin{equation}\label{unc2a}
\triangle X(t)\;=\;{1\over 2} \left[1+C(t) \left(\mathrm{e}^{-\mu
t}\tanh \varrho_{0} -\cos \left(\theta_{0} -2\omega t\right)
\right)\right]^{1/2}\,,
\end{equation}
\begin{equation}\label{unc2b}
\triangle Y(t)\;=\;{1\over 2} \left[1+ C(t) \left(\mathrm{e}^{-\mu
t}\tanh \varrho_{0} +\cos \left(\theta_{0} -2\omega
t\right)\right)\right]^{1/2}\,,
\end{equation}
where
$$
C(t)\;=\;\frac{2\mathrm{e}^{-\mu t}\tanh \varrho_{0}}{1-
\mathrm{e}^{-2\mu t}\tanh ^{2}\varrho_{0}} \,.
$$
The squeezing of $X$ occurs for
\begin{equation}\label{xsqu}
\cos \left(\theta_{0} -2\omega t\right)
> \mathrm{e}^{-\mu t}\tanh \varrho_{0}\,,
\end{equation}
while the squeezing of $Y$ occurs for
\begin{equation}\label{ysqu}
\cos \left(\theta_{0} -2\omega t\right)
 < -\mathrm{e}^{-\mu t}\tanh \varrho_{0}\,.
\end{equation}

\begin{figure}
\begin{center}
\includegraphics[width=4.8cm,height=4.8cm,angle=270]{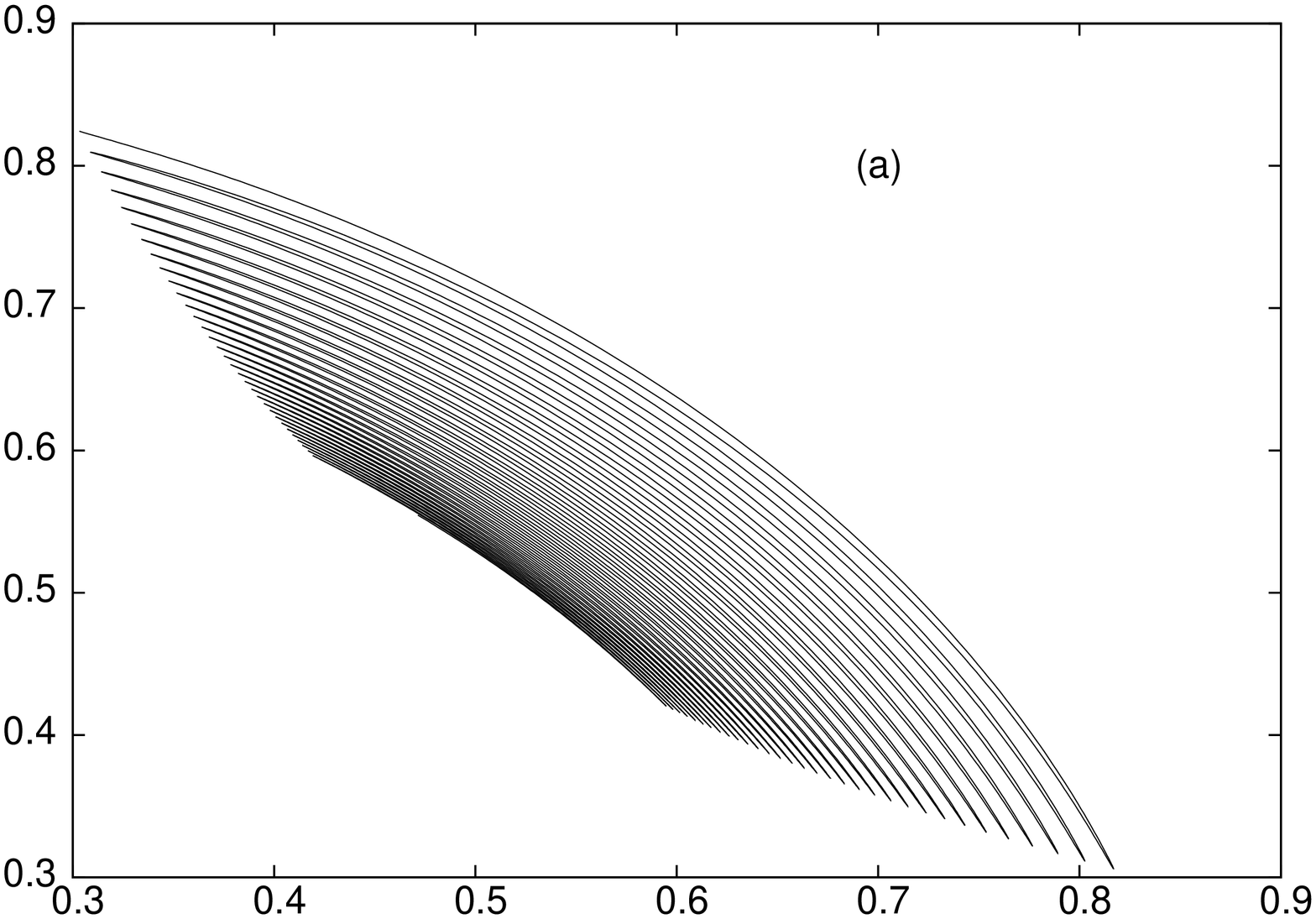}\\
\includegraphics[width=4.8cm,height=4.8cm,angle=270]{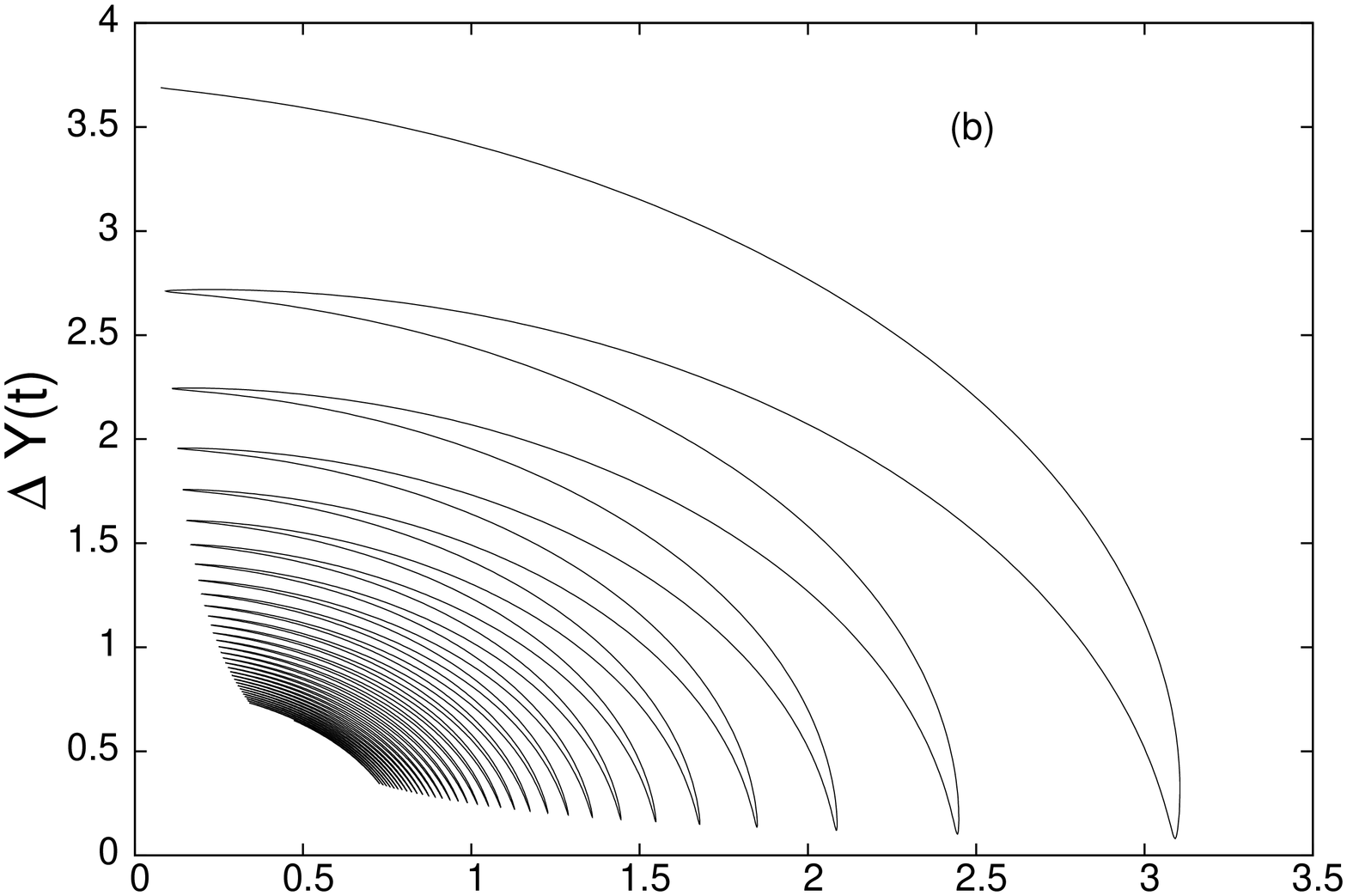}\\
\vspace*{.1ex}
\includegraphics[width=4.65cm,height=4.7cm,angle=270]{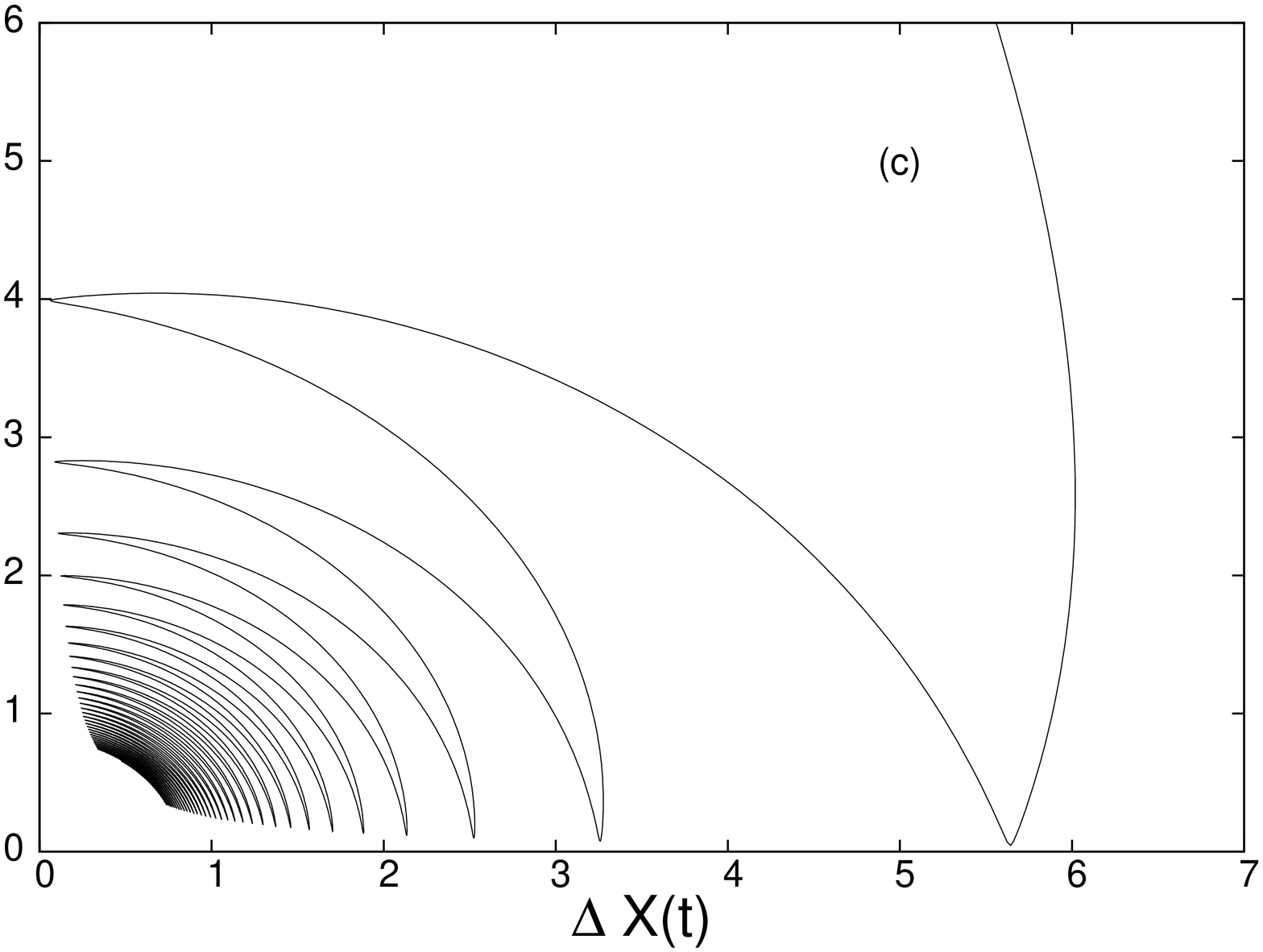}
\caption{Time dependence of the uncertainties $\Delta X$ and
$\Delta Y$ as given by Eqs. (\ref{unc2a}-\ref{unc2b}). The
dependence of $\Delta X$ and $\Delta Y$ on the dimensionless time
$\tau = \omega t$ is displayed for $\mu = 0.01 \omega$,
$\theta_{0} = 0$, $\vartheta=0.05$ and for three values of $\rho_0$: $0.5$ (a),
$2.0$ (b), and $8.0$ (c).}
\end{center}
\end{figure}

The time dependence of the uncertainties $\Delta X$ and $\Delta Y$
has been illustrated by the parametric plots presented in Fig.~1.
They show the dynamics of $\Delta X$ and $\Delta Y$ as functions
of the dimensionless time $\tau = \omega t$ ($0 \leq \tau \leq
100$) for $\mu = 0.01 \omega$ $\theta_{0} = 0$, and three values
of $\rho_{0}$. The pictures for different values of $\rho_{0}$
differ mostly for short times, otherwise they are qualitatively
very similar. From the above formulae as well as from the figure
it is clear that the dynamics of quadrature uncertainties is such
that the system switches back and forth from being squeezed in one
of the quadratures but not in the other while passing also through
the region where there is no squeezing at all. Asymptotically, the
system approaches the vacuum state (with $\Delta X = \Delta Y =
\frac{1}{2}$). An interesting feature of the envelopes of the
displayed curves is that the region where they are concave shrinks
as time grows, and the envelopes finally become convex.

A similar result concerning the uncertainties in the posterior
optical quadratures as well as the squeezing coefficients was
obtained in \cite{GoeGra94} where subharmonic generation from the
vacuum was studied: the uncertainties and $\eta(t)$ were found to
be deterministic, while $\alpha(t)$ was generated from the output
noise.

It may be appropriate at this point to recall that the posterior
wave function satisfying the  linear version of the filtering
equation is normalized to the probability density of the output
(observed) diffusion process with respect to the standard Wiener
measure of the input Wiener diffusion process. In the considered
case of the double heterodyne observation Eq.~(\ref{squ2}) implies
$\| \widehat{\psi}(t) \|^2=|l(t)|^2 $, therefore $|l(t)|^2$ is the
probability density of the complex Wiener process $\widehat{Q}(t)$
with respect to the standard Wiener measure of the complex input
process $Q(t)$. Note that the function $l(t)$ is essential in the
considered continuous observation. For example, as one can see from
(\ref{sol2}) that $l(t)$ depends on all the parameters of an
initial squeezed coherent state. If the values of some of them or
even all the values of these parameters are unknown, they can be
determined form $l(t)$.

We have proved that the filtering equation makes it possible to
study the time-development of a squeezed coherent state. This would
not be possible with the help of the master equation which can be
obtained from the filtering equation for the mixed posterior state
obtained from the Eq.~(\ref{filtequ1}) by taking the stochastic
average over all possible trajectories of the observed process.
Though the master equation (of the same form for both cases of the
heterodyne observation considered in the paper) preserves an
initially coherent state \cite{WalMil85}, it does not preserve the
squeezed coherent one \cite{ADabPhD}.

\section{Single balance heterodyne detection}

The filtering equation for a single balance heterodyne detection
\begin{eqnarray}\label{filtequ4}
&& \!\!\!\!\!\!\!\!\!\!\! \mathrm{d}\widehat{\psi}(t)=
-\bigg(\frac{\mathrm{i}}{\hbar}H+
\frac{\mu}{2}a^{\dagger}a\bigg)\widehat{\psi}(t)
\mathrm{d}t+\sqrt{\mu}\, a\,\mathrm{e}^{-\mathrm{i}\phi(t)}
\widehat{\psi}(t) \, \mathrm{d}\mathcal{Q}(t)\,,\nonumber
\\ && \!\!\!\!\!\!\!\!\!\!\!  \widehat{\psi}(0)=\psi\,,
\end{eqnarray}
can be obtained from Eq. (\ref{filequ}) by putting $\mu_{1}=2\mu$,
$\mu_{2}=0$, $\phi_{1}(t)=\phi(t)$,
$\mathcal{Q}_{1}(t)=\mathcal{Q}(t)$.

We shall prove that the squeezed coherent state is preserved under
the diffusion observation. For this purpose we employ the property
\begin{equation}\label{eigen2a}
S(\xi )a\,S^{\dagger}(\xi )|\xi ,\alpha \rangle \;=\;\alpha\, |\xi
, \alpha \rangle\,.
\end{equation}
Making use of the Baker-Hausdorff formula one can get the unitary
transform of the operator $a$ \cite{WalMil94}
\begin{equation}\label{eigen2b}
S(\xi )aS^{\dagger}(\xi )\;=\;a\,\Gamma _{1}+a^{\dagger}\,\Gamma _{2}\,,
\end{equation}
where $\Gamma _{1}=\cosh \varrho$,
$\Gamma_{2}=\mathrm{e}^{\mathrm{i} \theta} \sinh \varrho$.

Let us notice that if the system remains in the squeezed coherent
state (\ref{squ2}) at any time instant  $t\geq 0$, then the
following relations have to be satisfied
\begin{equation}\label{eigen3a}
S(\xi(t) )aS^{\dagger}(\xi(t) )\widehat{\psi}(t)\;=\;\left[a\, \Gamma _{1}
(t)+a^{\dagger}\,\Gamma _{2}(t)\right]\widehat{\psi}(t)\,,
\end{equation}
\begin{eqnarray}\label{eigen3b}
\lefteqn{S(\xi(t+\mathrm{d}t) )aS^{\dagger}(\xi(t+\mathrm{d}t)
)\widehat{\psi} (t+\mathrm{d}t)} \nonumber \\
&& = \left[a\, \Gamma _{1}(t+\mathrm{d}t)+a^{\dagger}\, \Gamma
_{2}(t+\mathrm{d}t)\right]\widehat{\psi}(t+\mathrm{d}t)\,.
\end{eqnarray}
Eqs. (\ref{eigen3a}) and (\ref{eigen3b}) can be reduced to the
single condition
\begin{eqnarray}\label{eigen3c}
&&\!\!\!\!\!\!\!\!\left[a \left(\Gamma
_{1}(t)\!+\!\mathrm{d}\Gamma _{1}(t) \right)\!+\! a^{\dagger}\,
\left( \Gamma _{2}(t)\!+\!\mathrm{d}\Gamma _{2}(t)\right)
\!-\!\alpha (t)\!-\!\mathrm{d}\alpha(t)\right] \mathrm{d}\widehat
\psi
(t)\nonumber\\
&&\!\!\!\!+\left(a\,\mathrm{d}\Gamma
_{1}(t)+a^{\dagger}\,\mathrm{d}\Gamma _{2}(t)-\mathrm{d}\alpha
(t)\right)\widehat \psi (t)\;=\;0\,.
\end{eqnarray}
Then by insertion of $\mathrm{d}\widehat{\psi}(t)$ from
Eq.~(\ref{filtequ4}) into Eq.~(\ref{eigen3c}) we obtain the set of
the differential equations
\begin{eqnarray}\label{diff3a}
&& \!\!\!\!\!\!\!\!\!\!\!\! \alpha(t)\Gamma_{1}(t)\!\left[-\Gamma_{1}(t)
\left(\mathrm{i}\omega \!+\!
\frac{\mu}{2}\right)\mathrm{d}t\!+\!\mathrm{d}\Gamma_{1}(t)\!+\!\! \mu\,
\mathrm{e}^{-2\mathrm{i}\phi(t)}\Gamma_{2}(t)\mathrm{d}t\right]
\nonumber\\
&&\!\!\!\!\!\!\!\!\!-\mathrm{d}\alpha(t)
\!-\!\alpha(t)\,\overline{\Gamma}_{2}(t)\left[\Gamma_{2}(t) \left(
\mathrm{i}\omega\!+\!\frac{\mu}{2}\right)\mathrm{d}t\!+\!
\mathrm{d}\Gamma_{2}(t)\right]\\
&&\!\!\!\!\!\!\!\!\!-\sqrt{\mu}\,\Gamma_{2}(t)\mathrm{e}^{-\mathrm{i}
\phi(t)}
\mathrm{d}\mathcal{Q}(t)\,=\,0\,,\nonumber
\end{eqnarray}
\begin{eqnarray}\label{diff3b}
&&\!\!\!\!\!\!\!\!\!\Gamma_{2}(t)\left[ \!-\mathrm{\Gamma}_{1}(t)
\left(\mathrm{i}\omega+
\frac{\mu}{2}\right)\mathrm{d}t+\mathrm{d}\mathrm{\Gamma}_{1}(t) +\!
\mu\, \mathrm{e}^{-2\mathrm{i}\phi(t)}\Gamma_{2}(t)\mathrm{d}t\right] \;
\nonumber\\
&&\!\!\!\!\!\!-\Gamma_{1}(t)\left[\mathrm{\Gamma_{2}(t)}\left(\mathrm{i}
\omega + \frac{\mu}{2}\right)\mathrm{d}t+
\mathrm{d}\mathrm{\Gamma}_{2}(t) \right]\;=\;0
\end{eqnarray}
with the initial condition: $\Gamma_{1}(0)=\cosh{\varrho_{0}}$,
$\Gamma_{2}(0)=\mathrm{e}^{\mathrm{i}\theta_{0}}\sinh
\varrho_{0}$, $\alpha(0)=\alpha_{0}$, and this completes the
proof.

The equation (\ref{diff3b}) imply that the function
$\mathrm{\Gamma}(t)=\mathrm{e}^{\mathrm{i}\theta(t)}\tanh \varrho(t)$
satisfies the Riccati differential equation of the form
\begin{eqnarray}\label{Riccati}
&&\frac{\mathrm{d}}{\mathrm{d}t}\mathrm{\Gamma}(t)\;=\;
-2\left(\mathrm{i} \omega+\frac{\mu}{2}\right)\mathrm{\Gamma}(t)+
\mu\,
\mathrm{e}^{-2\mathrm{i}\phi(t)}\,\mathrm{\Gamma}^{2}(t)\,,\nonumber
\\ &&
\Gamma(0)\;=\;\mathrm{e}^{\mathrm{i}\theta_{0}}\tanh\varrho_{0}\,.
\end{eqnarray}
Hence the posterior uncertainties of quadratures for the posterior
squeezed coherent state, given by the formulae
\begin{equation}
\label{deltax2}
\Delta X(t)\;=\; \left(4 \mathrm{Re} \kappa(t)\right)^{-1/2}\,,
\end{equation}
\begin{equation}
\label{deltay2}
\Delta Y(t)\;=\;|\kappa(t)|\left(4 \mathrm{Re} \kappa(t)\right)^{-1/2}\,,
\end{equation}
where
\begin{equation}
\kappa(t)\;=\; \frac{1+\mathrm{\Gamma}(t)}{1-\mathrm{\Gamma}(t)}
\end{equation}
do not depend on the measured noise, as before. The general solution to
Eq. (\ref{Riccati}) can be written as
\begin{equation}
\mathrm{\Gamma}(t)\;=\; \frac{\mathrm{\Gamma}(0)\mathrm{e}^{-(2\mathrm{i}
\omega+\mu)t}}{1 -\mu\, \Gamma(0)\int\limits_{0}^{t}
\mathrm{e}^{-\big(2\mathrm{i}
\omega+\mu\big)t^{\prime}-2\mathrm{i}\phi(t^{\prime})}\mathrm{d}
t^{\prime}}\,.
\end{equation}
In particular, for the phase $\phi(t)=\pi/2+\vartheta t $, we obtain
\begin{equation}
\mathrm{\Gamma}(t)\;=\; \frac{\big(2\mathrm{i}\omega+2\mathrm{i}
\vartheta+
\mu\big)\mathrm{\Gamma}(0)}{ \mathrm{e}^{(2\mathrm{i}\omega+\mu)t}
\big[2\mathrm{i}\omega+2\mathrm{i}\vartheta+ \mu(1+\Gamma(0))\big]-\mu\,
\Gamma(0)\mathrm{e}^{-2\mathrm{i}\vartheta t}}\,.
\end{equation}

The time dependence of the uncertainties $\Delta X$ and $\Delta Y$
computed above has been illustrated
in Fig.~2 showing the dynamics of $\Delta X$ and $\Delta Y$ (with
the same values of all parameters as in Section~3).
It is very hard to find any qualitative difference between the
corresponding parts of Fig.~2 and Fig.~1. Such a difference exists
only for small times. This is because the crucial variable
$\Gamma(t)$ becomes asymptotically exponential for large times,
that is, it behaves in the same way as in the case considered in
Section~3.

\begin{figure}
\begin{center}
\includegraphics[width=4.8cm,height=4.8cm,angle=270]{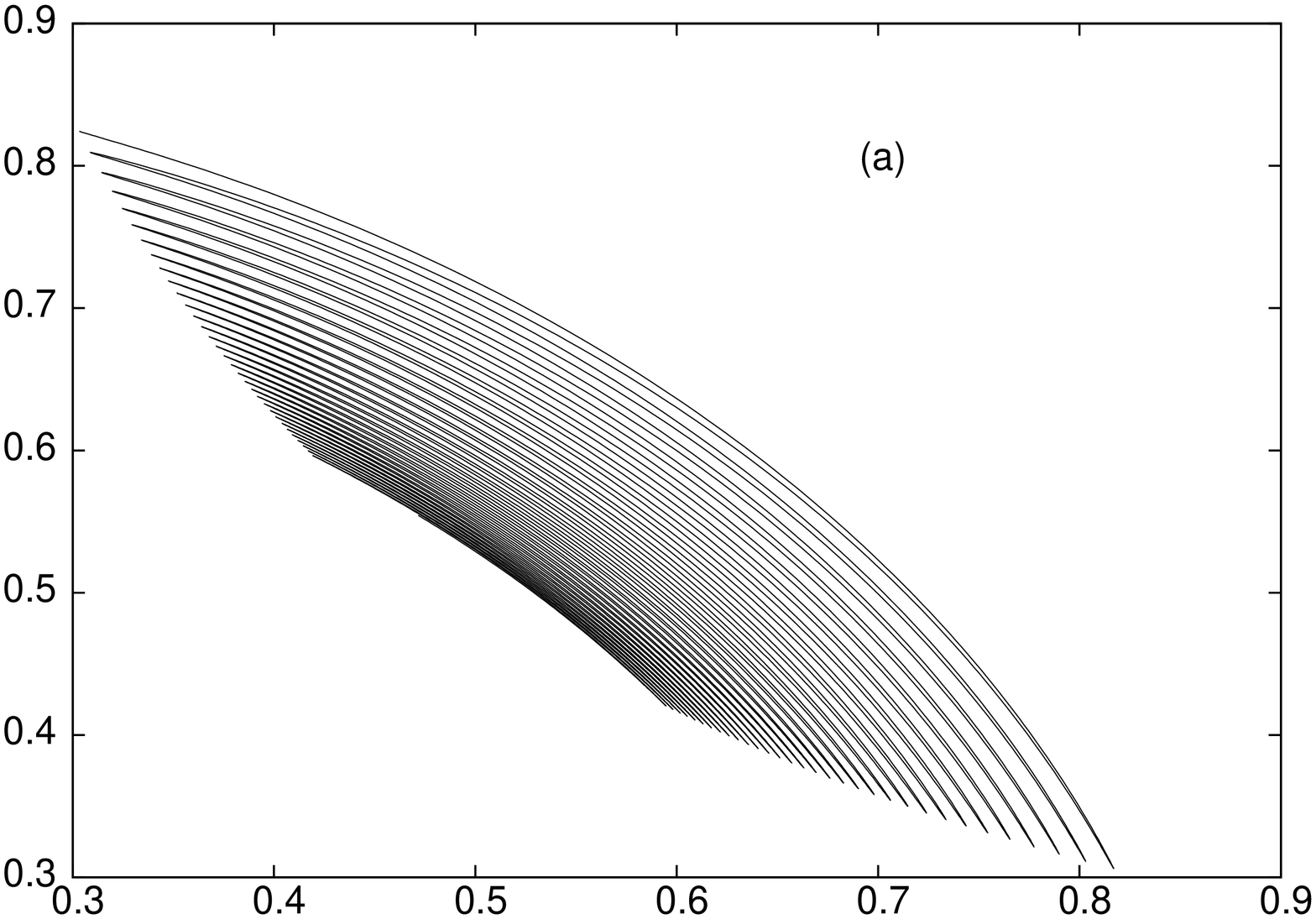}
\includegraphics[width=4.8cm,height=4.8cm,angle=270]{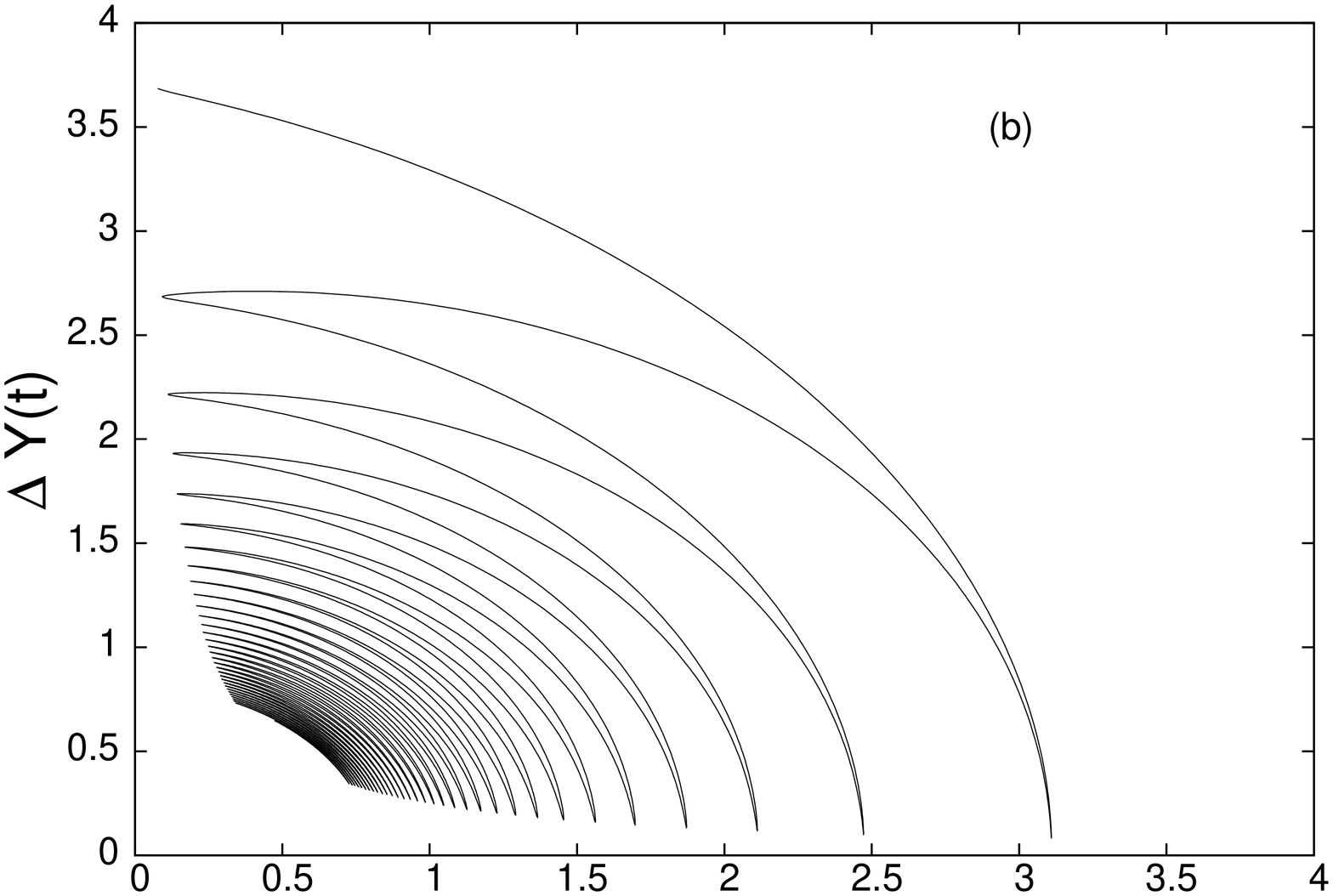}
\vspace*{.1ex}
\includegraphics[width=4.8cm,height=4.8cm,angle=270]{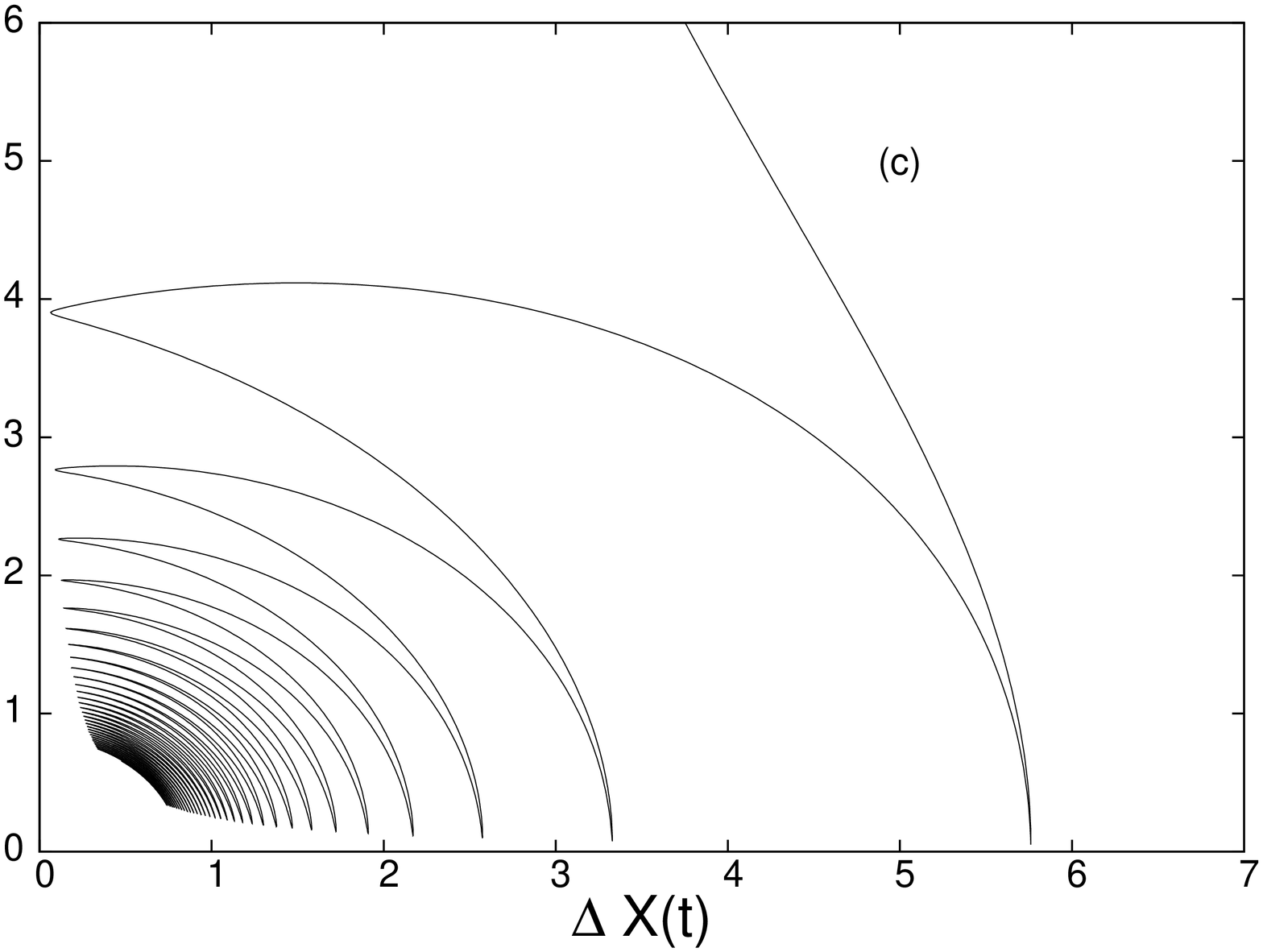}
\caption{Same as Fig.~1 except for Eqs.~(36-37) with
$\phi(t)=\pi/2+\vartheta t $.}
\end{center}
\end{figure}
To derive the differential equation for the coefficient $l(t)$ one
has to insert the state $l(t)|\xi(t), \alpha(t)\rangle$ to the
filtering equation  (\ref{filtequ4}). The calculations are more
complicated in comparison with these of Section~3. Finally one
gets:

\begin{eqnarray}\label{solforel}
&&\!\!\!\!\!\!\!\!\!\!\!\!\!\!\!\!\!\!\!\!\!\!\!\!\! l(t)= \exp\bigg[
-\frac{\mathrm{i}\omega t}{2}+ \frac{1}{2}
(\left|\alpha(t)\right|^2-\left|\alpha_{0}\right|^2)\nonumber \\
&& \!\!\!\!\!\!\!\!\!\! -\mu\int_{0}^{t}\left(\frac{
\left|\Gamma(t)\right|^2}{2\left(1-\left|\Gamma(t)\right|^2\right)}-\frac{
\overline{\Gamma(t)}\alpha^{2}(t)}{1-\left|\Gamma(t)\right|^2}\right)
\mathrm{d}t \nonumber \\
&& \!\!\!\!\!\!\!\!\!\! -\frac{\mu}{2}\int_{0}^{t}\frac{
\mathrm{e}^{-2\mathrm{i}\phi(t)}}{1-\left|\Gamma(t)\right|^2}
\left( \frac{1}{2}\left|\Gamma(t)\right|^2\Gamma(t) +
\alpha^{2}(t)\right)\mathrm{d}t \nonumber \\
&& \!\!\!\!\!\!\!\!\!\! +\frac{\mu}{2} \int_{0}^{t}\frac{
(\overline{\Gamma(t)})^2\mathrm{e}^{2\mathrm{i} \phi(t)}}{
1-\left|\Gamma(t)\right|^2} \left(\frac{1}{2}\Gamma(t)-
\alpha^2(t)\right)\mathrm{d}t \\
&&  \!\!\!\!\!\!\!\!\!\! +\sqrt{\mu}\int_{0}^{t}
\frac{\mathrm{e}^{-\mathrm{i}\phi(t)}\alpha(t)}
{\sqrt{1-\left|\Gamma(t)\right|^2}}\,
\mathrm{d}\mathcal{Q}(t) \bigg] \nonumber
\end{eqnarray}
with
\begin{eqnarray}\label{solforalpha}
&&\!\!\!\!\!\!\!\!\!\!\!\! \alpha (t)= \frac{1}
{\sqrt{1-\left|\Gamma(t)\right|^2}}
\left[\alpha_{0} \sqrt{1-\left|\Gamma(0)\right|^2} \right. \nonumber \\
&&\!\!\!\!\!\!\!\!\! \left. \times \exp \left(-\mathrm{i}
\omega t -{\mu \over 2}
t+\mu\int_{0}^{t}\mathrm{e}^{-2\mathrm{i}\phi(t^{\prime})}
\Gamma(t^{\prime})\mathrm{d}t^{\prime}\right)
\right.\nonumber\\
&& \!\!\!\!\!\!\!\!\! -\sqrt {\mu}\,\int_{0}^{t}
\exp\left(-\left(\mathrm{i}
\omega+\frac{\mu}{2}\right)(t-s) \right.\\
&& \!\!\!\!\!\!\!\!\!\left.\left.+ \mu\int_{s}^{t}
\mathrm{e}^{-2\mathrm{i}\phi(t^{\prime})}
\Gamma(t^{\prime})\mathrm{d}t^{\prime}\right)
\mathrm{e}^{-\mathrm{i}\phi(s)}
\Gamma(s)\mathrm{d}\mathcal{Q}(s)\right]\,.\nonumber
\end{eqnarray}
\section{Final remarks}
We have shown that in contrast to the master equation, the
filtering equation describing the reduction of the quantum state
following the registered trajectory for a single and double
heterodyne measurement does not destroy the squeezed coherent state
(\ref{squ2}). We have checked that for the initial state $|\xi_{0},
\alpha_{0}\rangle$ with $\xi_{0}\neq 0$, the amount of squeezing
decreases in time and registering the trajectory increases our
knowledge about $\mathcal{S}$. It is also worth to emphasize that
for the system prepared initially in a coherent state, the prior
and posterior mean values of the system operators coincide. The
posterior coherent state has a random phase and decreasing in time
independent of the noise amplitude. Even this case one can take
advantage from using the filtering equation: the probability
density $|l(t)|^2$ depends on $\alpha_0$, so $l(t)$ gives
information on the initial state of the oscillator. The analytical
solutions to the filtering equation for the initial coherent state
obtained for the diffusion observation in \cite{BarBel91, GoeGra94,
GoeGraHaa95}, and for the counting process in \cite{Car93} are
consistent with our results. Finally, let us note that it will be
very interesting to consider the posterior evolution of a squeezed
coherent state when a driving force is applied to the oscillator.

\end{document}